\documentclass[english,showkeys,twocolumn,superscriptaddress, showpacs]{revtex4-1}
\usepackage[T1]{fontenc}
\usepackage[latin9]{inputenc}
\usepackage{amsmath}
\usepackage{amssymb}
\usepackage{graphicx}

\makeatletter
\newenvironment{lyxlist}[1]
	{\begin{list}{}
		{\settowidth{\labelwidth}{#1}
		 \setlength{\leftmargin}{\labelwidth}
		 \addtolength{\leftmargin}{\labelsep}
		 }}
	{\end{list}}

\usepackage[colorlinks,
            linkcolor=blue,
            anchorcolor=blue,
            citecolor=blue
            ]{hyperref}

\makeatother

\usepackage{babel}

\begin{document}
\title{Effects of postselected von Neumann measurement on nonclassicality
of single-photon- added coherent state}
\author{Yusuf Turek}
\email{yusufu1984@hotmail.com}

\affiliation{School of Physics and Electronic Engineering, Xinjiang Normal University,
Urumqi, Xinjiang 830054, China}
\date{\today}
\begin{abstract}
The effects of von Neumann postselected measurement on  nonclassicality
of single-photon-added coherent state (SPACS) are studied. Explicit
expressions and analytical results for various field properties of
SPACS such as the photon number distribution, the Mandel $Q_{m}$
factor and the squeezing parameter of field quadrature after postselected
von Neumann measurement are investigated. The results showed that the
nonclassicality of SPACS after measurement changed dramatically than
initial state. The measurement let SPACS possess more strong sub-Poissonian
photon statistics in some definite coupling strength regimes and large
weak values which accompanied by low postselection probabilities.
\end{abstract}
\maketitle

\section{Introduction }

The preparation and optimization of non-classical quantum states have
great importance in quantum information processing including single
photon generation and detection \citep{Buller2010}, gravitational
wave detection \citep{Khalili2009,Grote2016}, quantum teleportation
\citep{Enk2001,Jeong2001,Milburn1999,Braunstien1998}, quantum computation
\citep{Ralph2003}, generation and manipulation of atom-light entanglement
\citep{Dudin2013,Hacker2019,Muschik2006}, and precision measurement
\citep{Munro2002} etc.. It is well known that the implementation
of those processes depends on the generation and optimization of the
related input radiation fields such as coherent states \citep{Glauber1963,Glauber1966,Stoler1971},
squeezed states \citep{Walls1983,Carranza2012,Andersen2016}, even
and odd coherent states \citep{Monroe1996,Ourjoumtsev2007,Yuen1976,Dodonov1995},
displaced and squeezed number states \citep{Yuen1976-1}, and binomial
states \citep{Stoler1985,Lee1985}. Thus, it is worthy to study the
inherent properties of radiation fields to find the suitable quantum
states for effectively implement the above mentioned quantum information
processing.There are some radiation fields which initially have classical
properties, but after added some photons possess nonclassical properties.
The photon- added coherent states (PACSs) are typical example.

PACSs was introduced by Agarwal and Tara in 1991\citep{Agarwal1991},
and this state exhibits an intermediate property between a classical
coherent state and a purely quantum Fock state. If we only consider
the one photon excitation of a classical coherent field, the generated
state is called single-photon-added coherent state (SPACS). SPACS
not only have interesting properties, but also be useful for possible
future applications including the engineering of quantum states \citep{Lund2004}
and quantum information protocols \citep{Wenger2004}, entangle state
generation \citep{Jing2017}. After Zavatta et al first studied the
experimental generation of the SPACS and visualized the evolution
of the quantum-to-classical transition \citep{Zavatta2004}, various
schemes to generation \citep{Jing2008,Barbieri2010} and enhance the
nonclassicality of SPACS widely investigated \citep{Dodonov1998,Zavatta2005,Kalamidas2008}.
In general, the enhancement of nonclassicality of SPACS depends on
the optimization of this state, and the optimization of quantum states
are related to quantum measurement.

Measurement is a basic concept in physical science and any information
of the system can be obtained from the measurement processes. In general,
a measurement composed of three parts, i.e., measuring device, measured
system and environment. According the requirements of quantum measurement,
in a quantum measurement process, there should be have interaction
between measuring device (pointer) and measured system and this interaction
should be short enough to guarantee the measurement precision. Thus,
in quantum measurement theory interaction Hamiltonian can be expressed
by von Newmann Hamiltonian as $H=g\hat{A}\otimes\hat{P}$ \citep{Nuemann1955}.
Here, $\hat{A}$ is the operator of measured system we want to measure,
$\hat{P}$ the canonical momentum of measuring device, and $g$ represent
the measuring strength. The measurement strength can decide the amount
of information of the measured system during after measurement. That
is to say, if the interaction strength between the measured system
and apparatus (measuring device) is strong, i.e. $g\gg1$, then we
can get the information of the system we want to get by single trial
with very small error. Whereas, if the interaction strength between
measured system and apparatus is very weak, i.e. $g\ll1$, the interference
between different eigenvalues of the system observable we want to
measure still exist and can't distinguish them. In weak coupling case
( $g\ll1$) we only can get very trivial information of the system
by single trial \citep{Aharonov1988}. However, the mission of getting
enough information of the system observable in weak coupling regime,
introduced a new kind of quantum measurement theory which called weak
measurement.

The weak measurement which characterized by postselection and weak
value was proposed by Aharonov, Albert, and Vaidman in 1988 \citep{Aharonov1988},
and considered as a generalized von Neumann quantum measurement theory.
In weak measurement theory, the coupling between the pointer and the
measured systems is sufficiently weak and the obtained information
by single trial is trivial. Even though the postselected weak measurement
on single system provides trivial information, by repeating it on
an arbitrarily large ensembles of identical system we can determine
the average result with arbitrary precision \citep{T2010} . One of
the distinguished properties of weak measurement compared with strong
measurement is that its induced weak value of the observable on the
measured system can be beyond the usual range of the eigenvalues of
that observable \citep{Aharonov2005}. The feature of weak value usually
referred to as an amplification effect for weak signals rather than
a conventional quantum measurement and used to amplify many weak but
useful information in physical systems. So far, the weak measurement
technique has been applied in different fields to investigate very
tiny effects, such as beam deflection \citep{Hosten2008,Turek2013,Pfeifer2011,Dixon2009,Starling2009},
frequency shifts \citep{Starling2010}, angular shifts \citep{Loaiza2014},
velocity shifts \citep{Viza2013}, and even temperature shift \citep{Egan2012}.
For details about the weak measurement and its applications in signal
amplification processes, we refer the reader to the recent overview
of the field \citep{Kofman2012,Dresel2014}. In weak measurement we
only consider the evolution of unitary operator upto its first order
since the interaction strength between the system and measuring device
very weak. However, if we want to connect the weak and strong measurement,
check to clear the measurement feedback of postselected weak measurement
and analyze experimental results obtained in nonideal measurements,
the full order evolution of unitary operator is needed \citep{Aharonov2005-1,Lorenzo2008,Pan2012},
we call this kind of measurement is postselected von Neumann measurement.

The signal amplification properties of weak measurement can be used
in state optimization problems. Recently, the state optimization problem
by using postselected von Neumann measurement have been presented
widely, such as taking the Gaussian states \citep{Kofman2012,Nakamura2012},
Hermite- Gaussian or Laguerre-Gaussian states, and non-classical states
\citep{Turek2015,Lima2014}. The advantages of non-classical pointer
states in increasing postselected measurement precision have been
examined in recent studies\citep{Turek2015,Pang2015,Turek2015-1}.
In Ref. \citep{Turek2018}, the authors studied the effects of postselected
measurement characterized by modular value \citep{Kedem2010} to show
the properties of semi-classical and non-classical pointer states
considering the coherent, coherent squeezed, and Schrodinger cat state
as a pointer. Most recently, the author of this paper investigated
the effects of postselected von Neumann measurement on the properties
of single-mode radiation fields \citep{Turek2019}, and found that
postselected von Neumann measurement really changed the photon statistics
and squeezing parameters of radiation fields for different weak values
and coupling strengths. However, to our knowledge, the effects of
postselected von Neumann measurement on the nonclassicality of SPACS
has not been previously investigated, neither exactly nor analytically,
in any literature.

In this paper, motivated by the previous studies \citep{Turek2018,Turek2019,Turek2015-1},
we investigate the effects of postselected von Neumann measurement
on nonclassicality of SPACS. In order to achieve our goal, we take
the spatial degree of freedom of SPACS as pointer and its polarization
degree of freedom as measured system. First of all, we obtained the
normalized final state of SPACS after postselected von Neumann measurement
by taking all interaction strengths between system and measuring device
into account. Then calculate the exact expressions and give numerical
results of physical quantities of SPACS such as photon number distribution,
$Q_{m}$factor and squeezing parameter. Thu numerical results showed
that postselected von Neumann measurement dramatically changed the
noncalssicality of SPACS. We notice that in our scheme the measurement
strengths and weak values have important role in changing the nonclassicality
of SPACS.

The rest of this paper is organized as follows. In Sec. \ref{sec:2},
we introduce the basic model setup for our scheme. In Sec. \ref{sec:3} and Sec. \ref{sec:4},
we give the details about the effects of postselected von Nuemann
measurement on the noncalssicality of SPACS. We calculate
the photon number distribution, Mandel $Q_{m}$factor and squeezing
parameter of SPACS, and found that postselected von Nuemann measurement
can change the nonclassicality of SPACS by changing the coupling strengths
and weak values. We give a conclusion to our paper in Sec. \ref{sec:5}.
Throughout this paper, we use the unit $\hbar=1$.

\section{\label{sec:2} model setup}

In this section, we introduce the related theories to our current
research. According to the standard quantum measurement theory, the
coupling interaction between system and measuring device is taken
to the standard von Neumann type Hamiltonian \citep{Nuemann1955}

\begin{equation}
H=g\delta(t-t_{0})\hat{A}\otimes\hat{P}.\label{eq:Hamil}
\end{equation}
Here, $g$ is a coupling constant and $\hat{P}$ is the conjugate
momentum operator to the position operator $\hat{X}$ of the measuring
device, i.e., $[\hat{X},\hat{P}]=i\hat{I}$. $\hat{A}$ is observable
of the measured system we want to measure. In general, to guarantee
the measurement precision, the interaction time between measured system
and measuring device is too short so that $\int_{0}^{T}g\delta(t-t_{0})dt=g$.
Thus, the time evolution operator $e^{-i\int_{0}^{T}Hdt}$ of our
system can be written as $e^{-igA\otimes\hat{P}}$.

The position operator $\hat{X}$ and momentum operator $\hat{P}$
of measuring device can be written in terms of annihilation (creation)
operators, $\hat{a}$($\hat{a}^{\dagger}$) in Fock space representation
as
\begin{eqnarray}
\hat{X} & = & \sigma(a^{\dagger}+a),\label{eq:annix}\\
\hat{P} & = & \frac{i}{2\sigma}(a^{\dagger}-a),\label{eq:anniy}
\end{eqnarray}
where $\sigma$ is the width of the beam. We know that the annihilation
(creation) operators $a$ ($a^{\dagger}$) obey the commutation relation,
$[a,a^{\dagger}]=1$. By substituting Eq.(\ref{eq:anniy}) to unitary
evolution operator $e^{-ig\hat{A}\otimes\hat{P}}$ for system observable
$\hat{A}$ which satisfies the property $\hat{A}^{2}=\hat{I}$, we
get
\begin{equation}
e^{-ig\hat{A}\otimes\hat{P}}=\frac{1}{2}(\hat{I}+\hat{A})\otimes D(\frac{s}{2})+\frac{1}{2}(\hat{I}-\hat{A})\otimes D(-\frac{s}{2}),\label{eq:UNA1}
\end{equation}
where parameter $s$ defined by $s:\equiv g/\sigma$ and it can characterize
the measurement strength of our scheme, and $D(\alpha)$ is a displacement
operator with complex $\alpha$ defined by $D(\alpha)=e^{\alpha a^{\dagger}-\alpha^{\ast}a}$
. Note that we can say that the coupling between system and pointer
is weak (strong) if $s\ll1$$(s\gg1)$, and in this study we will
consider all interaction strengths between system and measuring device.

If we assume that initially the system prepared to $\vert\psi_{i}\rangle$
and the initial state of the measuring device is $\vert\Psi\rangle$,
then the evolution of the total system can be written as $e^{-ig\hat{A}\otimes\hat{P}}\vert\psi_{i}\rangle\vert\Psi\rangle$
. Since our mission is to study the measurement effects on the properties
of state $\vert\Psi\rangle$, the normalized final state of the measuring
device can be obtained by taking the postselection with postselected
state $\vert\psi_{f}\rangle$ onto $e^{-ig\hat{A}\otimes\hat{P}}\vert\psi_{i}\rangle\vert\Psi\rangle$.
After some calculation, it reads

\begin{equation}
\vert\Phi\rangle=\beta\left[(1+\text{\ensuremath{\langle\hat{A}\rangle_{w}}})D(\frac{s}{2})+(1-\langle\hat{A}\rangle_{w})D(-\frac{s}{2})\right]\vert\text{\ensuremath{\Psi}}\rangle.\label{eq:8}
\end{equation}
Here, the normalization coefficient $\beta$ defined by
\begin{align}
\beta & =\frac{1}{\sqrt{2}}[1+\vert\langle\hat{A}\rangle_{w}\vert^{2}+\gamma^{2}e^{-\frac{s^{2}}{2}}Re[(1+\langle A\rangle_{w})^{\ast}(1-\langle\hat{A}\rangle_{w})\text{\ensuremath{\times}}\nonumber \\
 & e^{2siIm[\alpha]}(1+(\alpha^{\ast}+s)(\alpha-s))]]^{-\frac{1}{2}},\label{eq:9}
\end{align}
and
\begin{equation}
\langle A\rangle_{w}=\frac{\langle\psi_{f}\vert\hat{A}\text{\ensuremath{\vert\psi_{i}\rangle}}}{\langle\psi_{f}\vert\psi_{i}\rangle}
\end{equation}
is the weak value of system observable $A$. In general, the weak
value $\langle A\rangle_{w}$ is complex and can beyond the average
values of eigenvalues of observable $A$. That is to say, if the preselection
state $\vert\text{\ensuremath{\psi_{i}}}\rangle$ and postselection
state $\vert\psi_{f}\rangle$ are almost orthogonal, the $\langle A\rangle_{w}$
can take large values and this feature of weak measurement used as
amplification of weak signals of realated physical systems.

In this paper, we consider the spatial transversal freedom of SPACS
as measuring device and its polarization of freedom as measured system,
and study the effects of the measurement which performed on polarization
part on the inherent properties of spatial part of SPACS. Assume that
initially we prepared the system to the polarization state $\vert\psi_{i}\rangle=\cos\frac{\varphi}{2}\vert H\rangle+e^{i\delta}\sin\frac{\varphi}{2}\vert V\rangle$
and consider the observable $\hat{A}=\hat{\sigma}_{x}=\vert H\rangle\langle V\vert+\vert V\rangle\langle H\vert$.
After weak measurement, the system state postselected to $\vert\psi_{f}\rangle=\vert H\rangle$,
then the weak value $\text{\ensuremath{\langle\sigma_{x}\rangle_{w}}}$
can be obtained as
\begin{equation}
\langle\sigma_{x}\rangle_{w}=e^{i\delta}\tan\frac{\varphi}{2}.\label{eq:12}
\end{equation}
Here, $\vert H\rangle$ and $\vert V\rangle$ are horizontal and vertical
polarization of the beam with eigenvalues $1$ and $-1$, respectively.
From this expression we can see that this weak value is a complex
number and its value can be beyond the eigenvalues of $\hat{\sigma}_{x}$.
However, this large weak values accompanied by low successful postselection
probability, $P_{s}=\vert\langle\psi_{f}\vert\psi_{i}\rangle\vert^{2}=\cos^{2}\frac{\varphi}{2}$.

As mentioned earlier, in this paper, we consider SPACS as measuring
device, and its expression is given by
\begin{equation}
\vert\Psi\rangle=\gamma a^{\dagger}\vert\alpha\rangle\label{eq:7}
\end{equation}
where $\gamma=(1+\vert\alpha\vert^{2})^{-1}$ is normalization coefficient,
and $\vert\alpha\rangle=D(\alpha)\vert\alpha\rangle$ is coherent
sate with coherent parameter $\alpha=re^{i\theta}$. As proposed in
original paper of Agarwal \citep{Agarwal1991}, this state can be
produced in nonlinear processes in cavities, and its first implementation
is given with nonlinear optical scheme \citep{Zavatta2004}. It is
well known that if we subtract on photon from coherent radiation field,
the generated state still is a classical field. i.e., $a\vert\alpha\rangle=\alpha\vert\alpha\rangle$.
However, a single-photon excitation of a coherent state changes it
into something quite different. In other words, the application of
the creation operator $a^{\dagger}$ changes completely the classical
coherent state into a new quantum state which posses nonclassicality.
The purpose of our study is to check the effects of measurement on
the nonclassicality of SPACS by taking into account the signal amplification
property of postselected von Nuemann measurement.

\section{\label{sec:3} Effects on Photon Statistics}

In this section, in order to investigate the effects of postselected
von Nuemann measurement on photon statistics of SPACS, we check the
photon number distribution and Mandel $Q_{m}$ factor of SPACS after
measurement.

\subsection{Photon number distribution }

\begin{figure}
\includegraphics[width=4cm]{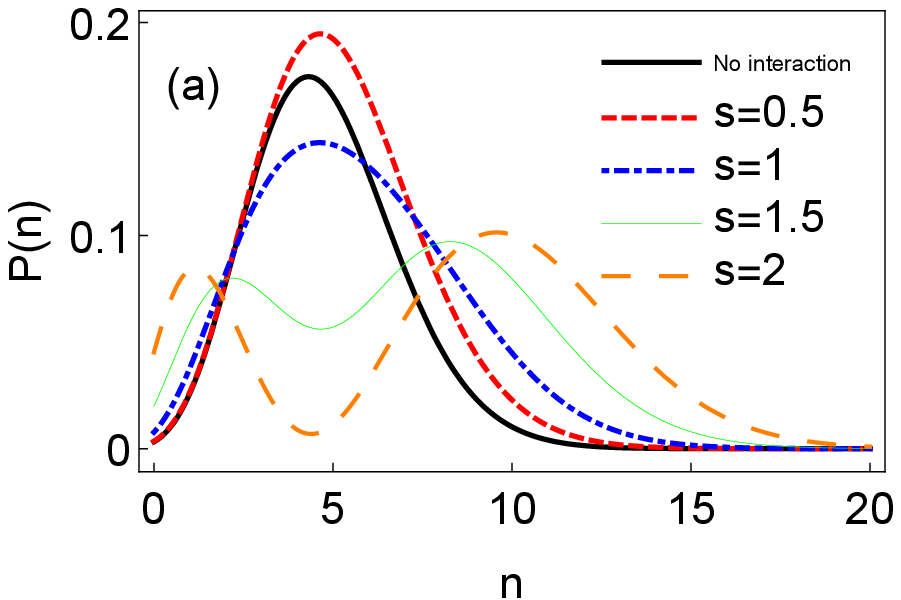}\includegraphics[width=4cm]{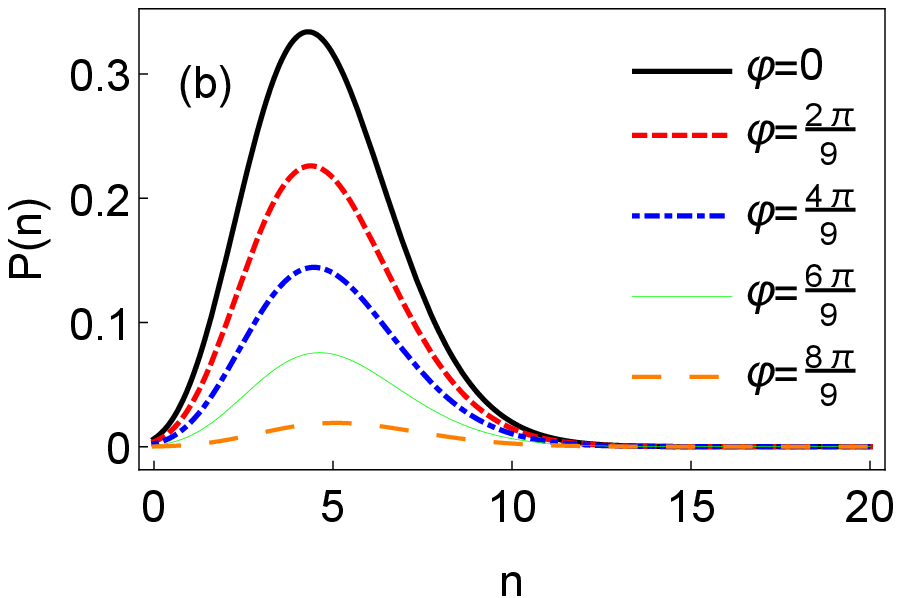}

\caption{(Color online) \label{fig:1} Photon number distribution $P(n)$ of
SPACS after postselected von Nuemann measurement as a function photon
numbers $n$ for different coupling strengths $s$ with fixed weak
value ($\varphi=\frac{\pi}{3}$) in (a), and for different weak values
with fixed coupling strength ($s=0.1$) in (b) . Here, $\delta=\frac{\pi}{4}$,
$r=2$, $\theta=\frac{\pi}{9}$. }
\end{figure}

In this subsection, we check the effect of postselected von Neumann
measurement on photon number distribution of SPACS. The probability
of finding $n$ photons under the final state $\vert\Phi\rangle$
of SPACS after measurement is given by
\begin{equation}
P(n)=\vert\langle n\vert\Phi\rangle\vert^{2}.
\end{equation}
 The explicit expression of $P(n)$ can be obtained by substituting
$\vert\Phi\rangle$ which is given in Eq. (\ref{eq:8}) to above expression,
and its corresponding numerical results are shown in Fig. \ref{fig:1}.
From previous studies it can be deduced that the initial SPACS have
sub-Poissonian distribution \citep{Agarwal1991}, and the solid
black curve in Fig. \ref{fig:1}(a)  represent the photon number distribution
of initial SPACS. In Fig. \ref{fig:1} (a) we check the effects of
interaction strengths on photon number distribution, the result showed
that with increasing the interaction strength, the photon number distribution
of SPACS become broader and occur oscillation in definite photon regions.
We know that weak value have amplification property on weak signals
when pre- and pos-selection of system state almost orthogonal. In
Fig.\ref{fig:1} (b) we display the photon number distribution $P(n)$
of SPACS as a function of photon numbers $n$ for different weak values.
As indicated in Fig. \ref{fig:1} (b), in weak measurement regime,
the probability of finding $n$ photons decreased as increases the
weak value, and photon distribution curves become narrower than initial
state. This implies that in weak measurement regime, after postcelect
von Neumann measurement, the sub-Poissonian property of SPACS is increased
with large weak values.

\subsection{the Mandel $Q_{m}$ factor of SPACS }

The photon variance of initial SPACS less than its mean photon number
so that it has nonclassicality which characterized by sub-Poissonian
photon statistics \citep{Agarwal1991}. We know that if the underling
$P$- function does not possess or negative Wigner function is negative
then the corresponding radiation field have nonclassicality, and SPACS
which defined in Eq. (\ref{eq:7}) possess both of them \citep{Agarwal1991}.
However, in this study, in order to check the effects of postselected
von Nuemann measurement on nonclassicality of SPACS, we choose the
experimentally accessible method---Mandel $Q_{m}$ factor. This parameter
is very useful to characterize the nonclassicality of any radiation
field and it was introduced by Mandel in 1979 \citep{Mandel1979}.
According to his study, if any distribution which is narrower than
Poissonian distribution must correspond to a nonclassical radiation
field, and this nonclassicality of that radiation field can be characterized
by Mandel $Q_{m}$ factor. The definition of Mandel $Q_{m}$ factor
is
\begin{equation}
Q_{m}=\frac{\langle(a^{\dagger}a)^{2}\rangle-\langle a^{\dagger}a\rangle^{2}-\langle a^{\dagger}a\rangle}{\langle a^{\dagger}a\rangle}.\label{eq:12-1}
\end{equation}
It is clear from the above expression that $Q_{m}=0$ for coherent
state and $Q_{m}=-1$ for Fock state. The negativity of $Q_{m}$ ($-1\ll Q_{m}<0$
) is a sufficient condition for the field to be nonclassical \citep{Agarwal2013}.
However, if $Q_{m}>0$ we can't make a conclusion about the nonclassicality
of the field.

\begin{figure}
\includegraphics[width=4cm]{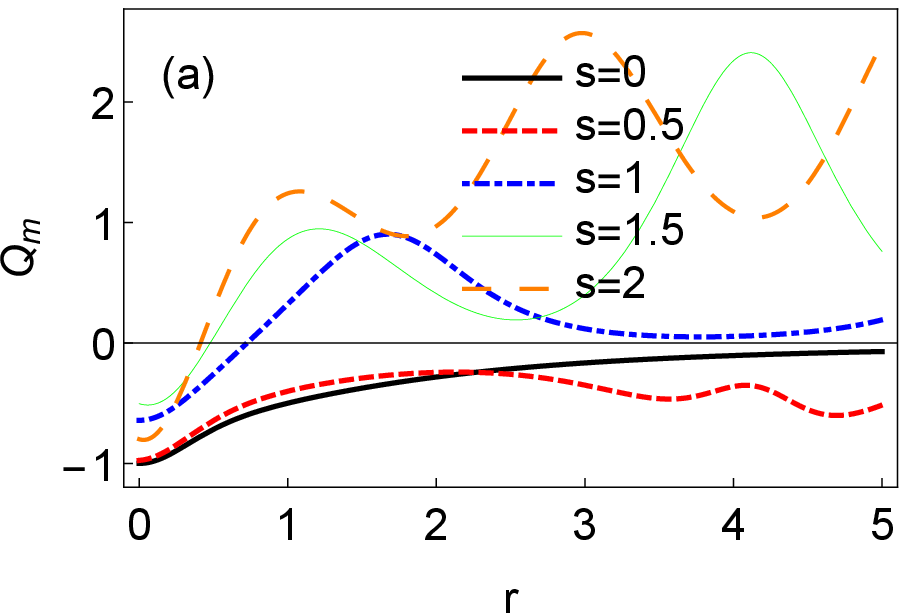}\includegraphics[width=4cm]{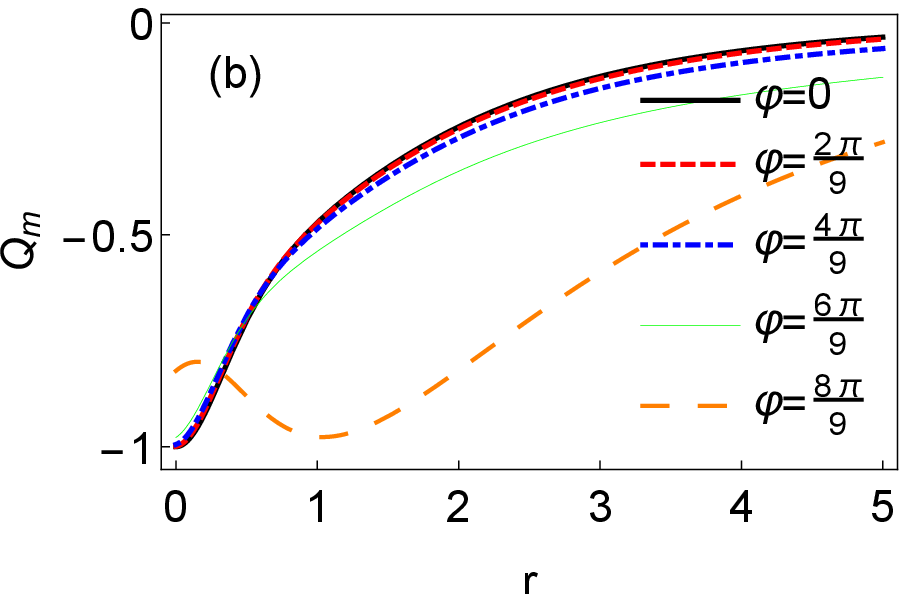}

\caption{(Color online) \label{fig:2} The Mandel $Q_{m}$ parameter of SPACS
after postselected von Neumann measurement as a function of coherent
state parameter $r$ for different coupling strengths $s$ with fixed
weak value ($\varphi=\frac{\pi}{9}$) in (a), and for different weak
values with fixed coupling strength ($s=0.1$) in (b) . Here, $\delta=0$,
$\theta=\frac{\pi}{4}$. }

\end{figure}

The $Q_{m}$ parameter for initial pointer state $\vert\Psi\rangle$
is given by
\begin{equation}
Q_{m,\Psi}=-\gamma^{2}\frac{1+2\vert\alpha\vert^{2}+2\vert\alpha\vert^{4}}{1+3\vert\alpha\vert^{2}+\vert\alpha\vert^{4}}.\label{eq:13-1}
\end{equation}
Furthermore, the $Q_{m,\Phi}$ can be obtained by calculating the
expectation values in Eq. (\ref{eq:12-1}) under the final normalized
state of the pointer $\vert\Phi\rangle$ after measurement. Since
the explicit expression of $Q_{m,\Phi}$ is too cumbersome to show
in this paper, we only give its numerical results in Fig. \ref{fig:2}.
In Fig.\ref{fig:2} (a), we plot $Q_{m,\Phi}$ as a function of coherent
state parameter $r$, and solid black curve ($s=0$) represent to
initial state case which is described by $Q_{m,\Psi}$ . It is showed
that the nonclassicality of photon added coherent state is attenuated
as coupling strength $s$ increases for definite weak value. In Fig.
\ref{fig:2} (b), we check the effects of different weak values on
Mandel $Q_{m,\Phi}$ parameter for fixed coupling strength. As indicated
in Fig.\ref{fig:2} (b), the nonclassicality of photon added coherent
state is increased as increases of weak value in weak coupling regime
($s<1$), and these results keep accordance with numerical results
of photon number distribution which presented in Fig. \ref{fig:1}.
 From the Fig. \ref{fig:2} (a) and (b) we can induce that in weak
coupling regime the photon statistics of photon added coherent state
gradually changed more sub-Poissonian as increases the weak value
which accompanied by low postselection probabilities.

\section{\label{sec:4} effects on squeezing parameter of the field}

In this section, we study the effects of postselected von neumann
measurement on squeezing parameter of SPACS. In general, the squeezing
parameter of radiation field is defined as \citep{Agarwal2013}
\begin{equation}
S_{\phi}=(\triangle X_{\phi})^{2}-\frac{1}{2}\label{eq:17}
\end{equation}
 where
\begin{equation}
X_{\phi}=\frac{1}{\sqrt{2}}(ae^{-i\phi}+a^{\dagger}e^{i\phi}),\ \ \phi\in[0,2\pi]\label{eq:18}
\end{equation}
is the quadrature operator of the field with
\begin{equation}
[\hat{X}_{\phi},\hat{X}_{\phi+\frac{\pi}{2}}]=i\label{19}
\end{equation}
 and
\begin{equation}
(\triangle X_{\phi})^{2}=\langle\Phi\vert X_{\phi}^{2}\vert\Phi\rangle-\langle\Phi\vert X_{\phi}\vert\Phi\rangle^{2}.\label{eq:20}
\end{equation}
 From the definition of $S_{\phi}$, it can be seen that $S_{\phi}\ge-\frac{1}{2}$.
If $-\frac{1}{2}\le S_{\phi}\le0$, then there have squeezing effect
of corresponding quadrature of radiation field. For initial pointer
state $\vert\Psi\rangle$, the squeezing parameter $S_{\phi,\Psi}$
is given by
\begin{equation}
S_{\phi,\Psi}=\gamma^{4}\left[1-\vert\alpha\vert^{2}\cos2\left(\phi-\theta\right)\right].\label{eq:21}
\end{equation}
 From this expression, Eq. (\ref{eq:21}), it can be deduced that
only when $\vert\alpha\vert^{2}>1$ with $\phi=\theta$, there will
occur a squeezing of the corresponding quadrature of SPACS. It is
well known that the coherent state is a minimum uncertainty state
and squeezing parameter is equal to zero. However, after added on
photon, its squeezing feature changed dramatically since the generated
new radiation field possess nonclassicality.
\begin{figure}
\begin{lyxlist}{00.00.0000}
\item [{\includegraphics[width=4cm]{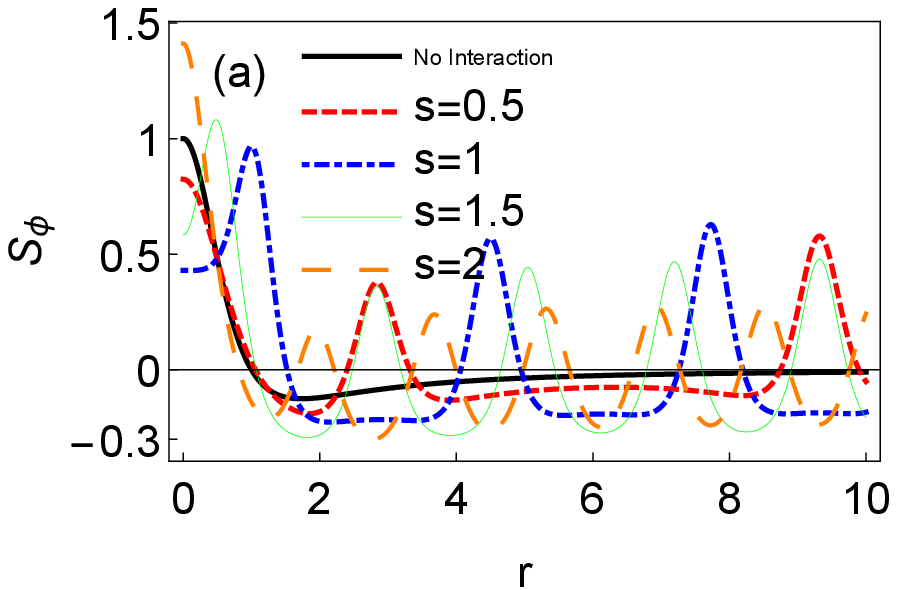}\includegraphics[width=4cm]{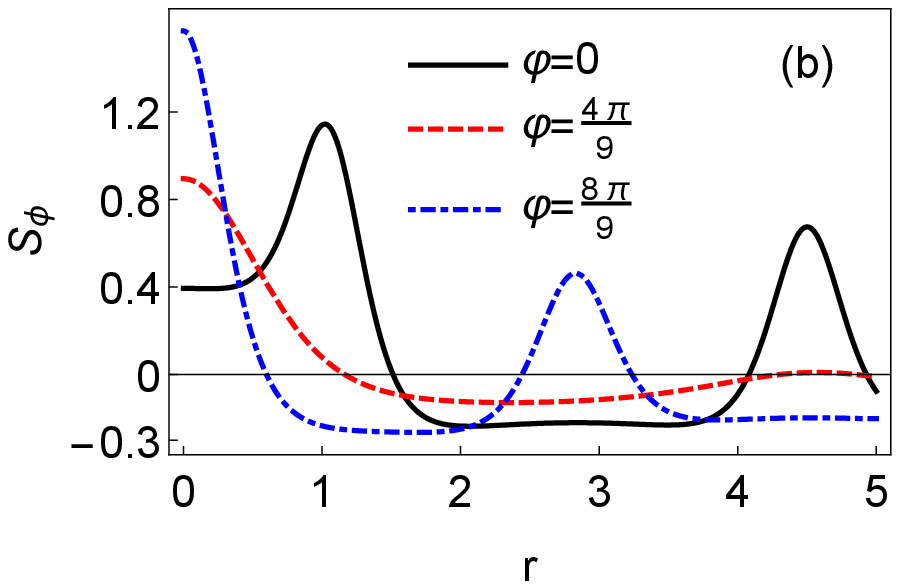}}]~
\end{lyxlist}
\includegraphics[width=4cm]{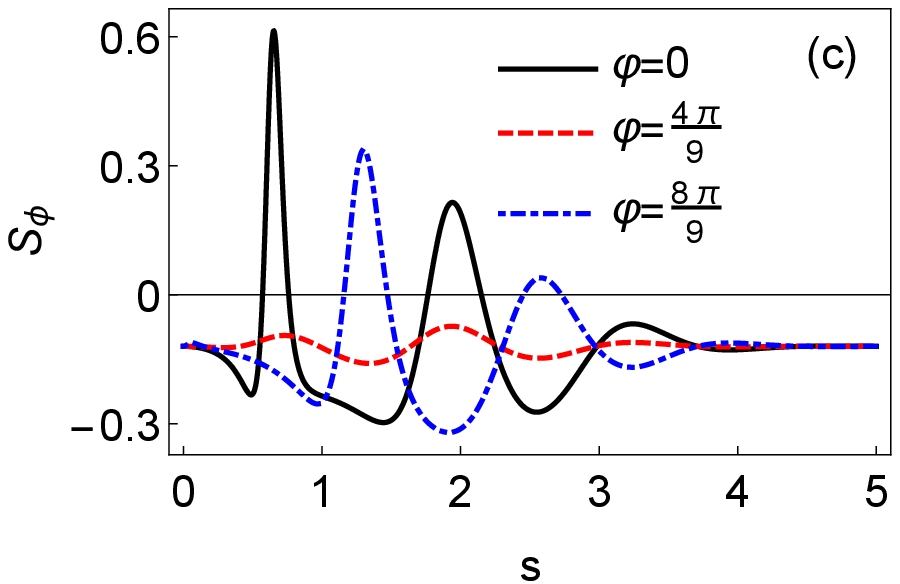}\includegraphics[width=4cm]{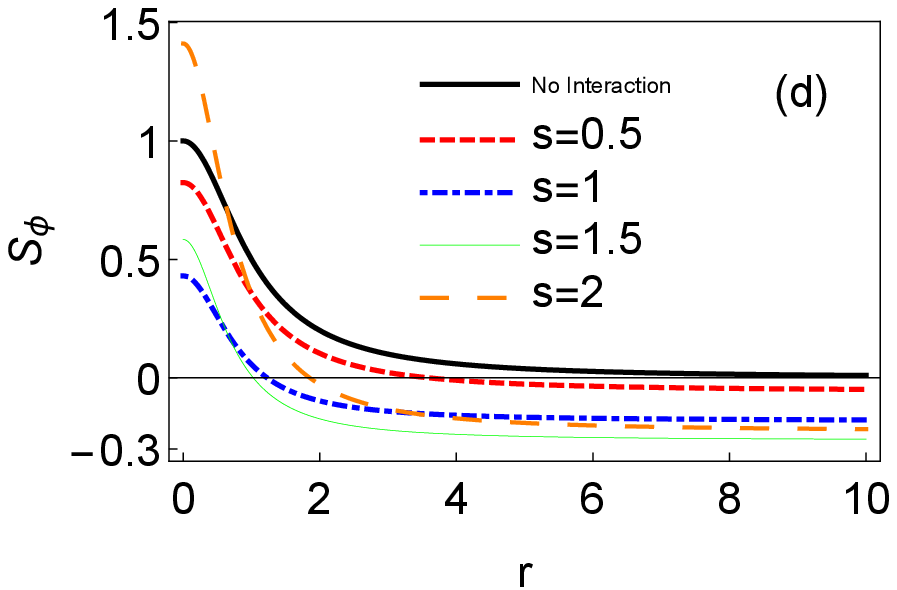}

\caption{\label{fig.3-1}(Color online) The squeezing parameter $S_{\phi}$
of SPACS after postselected von Nuemann measurement. Here,$\delta=0$,
and (a) $\varphi=\frac{\pi}{9}$ , $\theta=\frac{\pi}{2},$ $\phi=\frac{\pi}{2}$
; (b ) $s=1$, $\theta=\frac{\pi}{2},$ $\phi=\frac{\pi}{2}$ ; (c)
$r=2$, $\theta=\frac{\pi}{2},$ $\phi=\frac{\pi}{2}$ ; (d)  $\varphi=\frac{\pi}{9}$,
$\theta=0,$ $\phi=\frac{\pi}{2}$ . }
\end{figure}

\begin{figure}
\includegraphics[width=4cm]{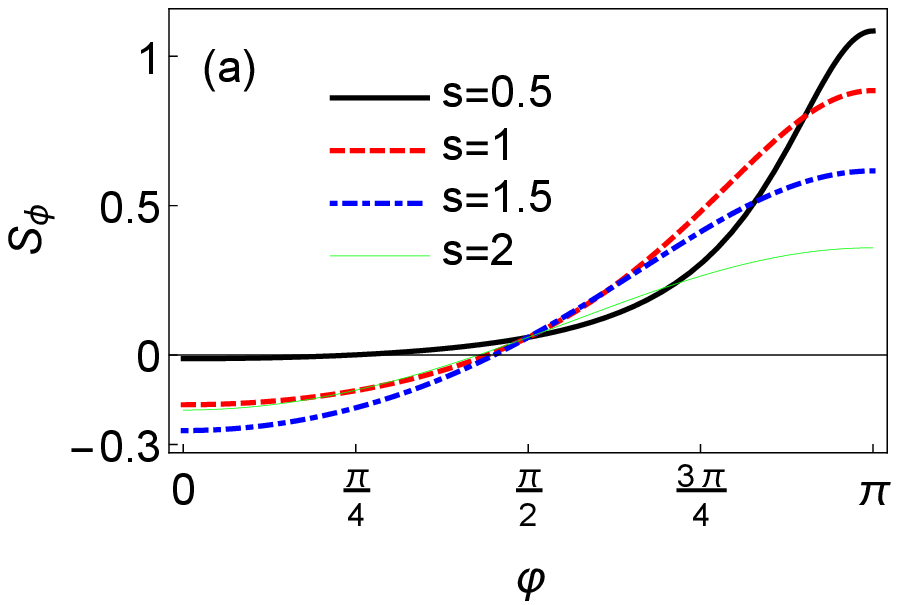}\includegraphics[width=4cm]{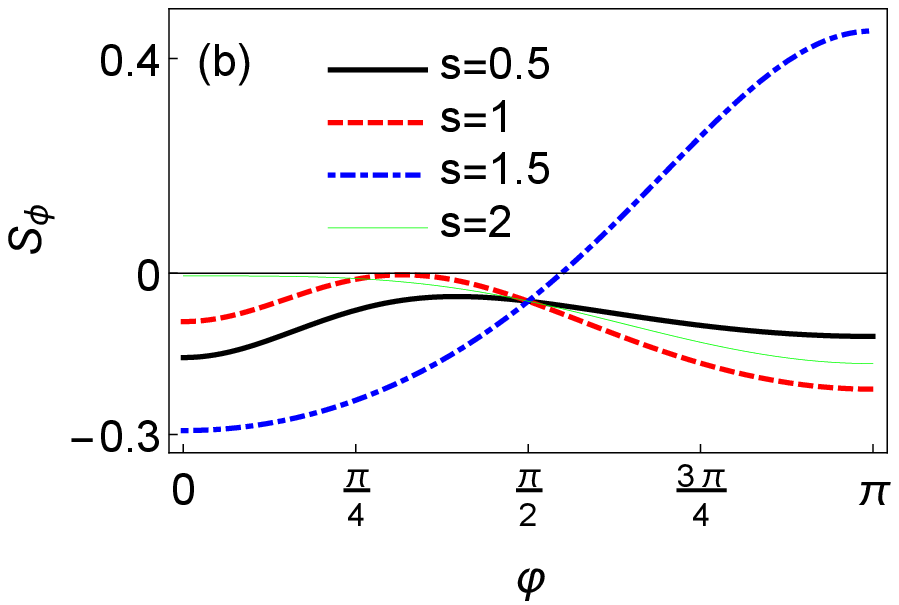}

\caption{\label{fig.4}(Color online) The squeezing parameter $S_{\phi}$ SPACS
after postselected von Nuemann measurement. (a) $\delta=0$, $r=4$,
$\theta=0,$ $\phi=\frac{\pi}{2}$; (b) $\delta=0$, $r=4$, $\theta=\frac{\pi}{2},$
$\phi=\frac{\pi}{2}$.}
\end{figure}

The explicit expression of squeezing parameter $S_{\phi}$ of SPACS
after postselected von neumann measurement can be calculated by using
the final state $\vert\Phi\rangle$ and Eqs. (\ref{eq:17}). Since
its exact expression is too cumbersome to show here, in this paper
we just analyzed the corresponding numerical results. As detailed
in Fig. \ref{fig.3-1} and Fig.\ref{fig.4}, the postselected von
nuemann measurement changed the squeezing effect of photon added coherent
state significantly. In Fig. \ref{fig.3-1}(a) we plot the $S_{\phi,\Phi}$
vs coherent state parameter $r$. The black solid curves in Fig. \ref{fig.3-1}
(a) represent the squeezing parameter of initial SPACS $\vert\text{\ensuremath{\Psi}}\rangle$
, i.g. $X_{\phi=\frac{\pi}{2},\Psi}$ , and other curves represents,
$S_{\phi=\frac{\pi}{2},\Phi}$, after postselected measurement with
various coupling coupling strengths for $X_{\phi=\frac{\pi}{2}}$
. It can be seen from Fig. \ref{fig.3-1}(a) that squeezing parameter
of SPACS is increased in postselected von nuemann measurement for
some regions. In Fig. \ref{fig.3-1}(b) and (c) we check the effects
of different weak values on squeezing effect of SPACS for fixed coupling
strength and fixed coherent state papramter $r$. The results showed
that weak values induced by postselected non Nuemann measurement can
increase the squeezing effect of SPACS for definite coupling strengths
compared to initial SPACS (see the black solid curve in Fig.\ref{fig.3-1}(a)).
As mentioned above, the squeezing effects of initial SPACS phase dependent
and if only the condition $\theta=\phi$ is satisfied then there will
occur the squeezing of corresponding quadrature of the SPACS. However,
as indicated in Fig. \ref{fig.3-1}(d), the condition $\theta=\phi$
is no more needed to realize the quadrature squeezing of SPACS after
postselected non Nuemann measurement. The result indicated that the
squeezing effect increased with increasing the coupling strength $s$
between system and measuring device compared to no interaction case
(see the black solid curve in Fig. \ref{fig.3-1} (d)).

In order to further study the effects of weak values on phase catch
condition of squeezing parameter of SPACS after poseslected von Nuemann
measurement, in Fig. \ref{fig.4} we plot the squeezing parameter
$S_{\phi=\frac{\pi}{2},\Phi}$ for various weak values. It is visible
by comparing the Figs. \ref{fig.4}(a) and (b) that the condition
$\theta=\phi$ is not necessary for realizing the squeezing effect
of field quadratures of SPACS after postselected von neumann measurement.

\section{\label{sec:5} Conclusion and remarks}

In this study, we investigated the effects of postselected von Nuemann
measurement on nonclassicality of SPACS. In order to achieve our goal,
first of all we drived explicit expression of SPACS after measurement
by considering all coupling strengths between system and measuring
device. We checked the photon number distribution, Mandel $Q_{m}$
factor and squeezing parameter of SPACS after postselected von Nuemann
measurement for various system parameters. Our numerical results are
given according to the exact expressions of corresponding physical
quantities and they showed that postselected von Nuemann measurement
can change the nonclassicality which characterized by sub-Poissonian
photon distribution of SPACS dramatically. We found that with increasing
the weak values the photon number distribution becomes narrower, the
negativity of Mandel $Q_{m}$ factor and squeezing parameter are increased
than initial state. Our numerical results also showed that after postselected
von Nuemann measurement, the squeezing parameter of SPACS is no more
suffer the rigid phase matching condition as initial state.

We anticipate that the presented optimization scheme of nonclassicality
of SPACS in this paper would be helpful to provide other effective
methods to implement the related practical problems in quantum information
processing as listed in introduction part. Furthermore, in this paper
we only consider one photon excitation of coherent radiation field,
and it is a simple case of PACSs. Thus, effects of postslected von
Nuemann measurement on radiation field properties of more that one
photon excitation of coherent radiation field is still worth to study.
Work along this line is in progress and results will be presented
in near future.
\begin{acknowledgments}
This work was supported by the National Natural Science Foundation
of China (Grant No. 11865017) and the Introduction Program of High
Level Talents of Xinjiang Ministry of Science.
\end{acknowledgments}


\begin{thebibliography}{10}
\bibitem{Buller2010} G. S. Buller and R. J. Collins, \href{https://doi/10.1088/0957-0233/21/1/012002}{\emph{Meas. Sci. Technol} \textbf{21}, 12002 (2010)}.

\bibitem{Grote2016} H. Grote, M. Weinert, R. X. Adhikari, C. Affeldt,
and H. Wittel, \href{https://doi/10.1364/OE.24.020107}{\emph{Opt. Express} \textbf{24}, 20107 (2016)}.

\bibitem{Khalili2009} F. Y. Khalili, H. Miao, and Y. Chen, \href{https://link.aps.org/doi/10.1103/PhysRevD.80.042006}{\emph{Phys. Rev. D} \textbf{80}, 042006 (2009)}.

\bibitem{Enk2001} S. J. van Enk and O. Hirota, \href{https://link.aps.org/doi/10.1103/PhysRevA.64.022313}{\emph{Phys. Rev. A} \textbf{64}, 022313 (2001)}.

\bibitem{Jeong2001} H. Jeong, M. S. Kim, and J. Lee, \href{https://link.aps.org/doi/10.1103/PhysRevA.64.052308}{\emph{Phys. Rev. A} \textbf{64}, 052308 (2001)}.

\bibitem{Milburn1999} G. J. Milburn and S. L. Braunstein, \href{https://link.aps.org/doi/10.1103/PhysRevA.60.937}{\emph{Phys. Rev. A} \textbf{60}, 937 (1999)}.

\bibitem{Braunstien1998} S. L. Braunstein and H. J. Kimble, \href{https://link.aps.org/doi/10.1103/PhysRevLett.80.869}{\emph{Phys. Rev. Lett.} \textbf{80}, 869 (1998)}.

\bibitem{Ralph2003} T. C. Ralph, A. Gilchrist, G. J. Milburn, W.
J. Munro, and S. Glancy, \href{https://link.aps.org/doi/10.1103/PhysRevA.64.042319}{\emph{Phys. Rev. A} \textbf{68}, 042319 (2003)}.

\bibitem{Dudin2013} L. Li, Y. O. Dudin, and A. Kuzmich, \href{https://doi.org/10.1038/nature12227}{ \emph{Nature} \textbf{498}, 466 (2013)}.

\bibitem{Hacker2019} B. Hacker, S.Welte, S. Daiss, A. Shaukat, S.
Ritter, L. Li, and G. Rempe, \href{https://doi.org/10.1038/s41566-018-0339-5}{\emph{Nat. Photon.} \textbf{13}, 110 (2019)}.

\bibitem{Muschik2006} C. A. Muschik, K. Hammerer, E. S. Polzik, and
J. I. Cirac, \href{https://link.aps.org/doi/10.1103/PhysRevA.73.062329}{\emph{Phys. Rev. A} \textbf{73}, 062329 (2006)}.

\bibitem{Munro2002} W. J. Munro, K. Nemoto, G. J. Milburn, and S.
L. Braunstein, \href{https://link.aps.org/doi/10.1103/PhysRevA.73.023819}{\emph{Phys. Rev. A} \textbf{66}, 023819 (2002)}.

\bibitem{Glauber1963} R. J. Glauber, \href{https://link.aps.org/doi/10.1103/PhysRev.131.2766}{\emph{Phys. Rev.} \textbf{131}, 2766 (1963)}.

\bibitem{Glauber1966} U. M. Titulaer and R. J. Glauber, \href{https://link.aps.org/doi/10.1103/PhysRev.145.1041}{\emph{Phys. Rev.} \textbf{145}, 1041 (1966)}.

\bibitem{Stoler1971} D. Stoler, \href{https://link.aps.org/doi/10.1103/PhysRevD.4.155}{\emph{Phys. Rev. D} \textbf{4}, 155 (1971)}.

\bibitem{Walls1983} Walls and F. D., \href{https://doi.org/10.1038/306141a0}{\emph{Nature} \textbf{306}, 141 (1983)}.

\bibitem{Carranza2012} R. Carranza and C. C. Gerry, \href{https://doi.org/10.1364/JOSAB.29.002581}{\emph{J. Opt. Soc. Am. B} \textbf{29}, 2581 (2012)}.

\bibitem{Andersen2016} U. L. Andersen, T. Gehring, C. Marquardt,
and G. Leuchs, \href{https://doi.org/10.1088/0031-8949/91/5/053001}{\emph{Phys. Scripta.} \textbf{91}, 053001 (2016)}.

\bibitem{Monroe1996} C. Monroe, D. M. Meekhof, B. E. King, and D.
J. Wineland, \href{https://doi.org/10.1126/science.272.5265.1131}{\emph{Science} \textbf{272}, 1131 (1996)}.

\bibitem{Ourjoumtsev2007} A. Ourjoumtsev, H. Jeong, R. Tualle-Brouri,
and P. Grangier, \href{https://doi.org/10.1038/nature06054}{\emph{Nature} \textbf{448}, 784 (2007)}.

\bibitem{Yuen1976} H. Yuen, \href{https://link.aps.org/doi/10.1103/PhysRevA.13.2226}{\emph{Phys. Rev. A} \textbf{13}, 2226 (1976)}.

\bibitem{Dodonov1995} V. V. Dodonov, V. I. Man¡¯Ko, and D. E. Nikonov,
\href{https://link.aps.org/doi/10.1103/PhysRevA.51.3328}{\emph{Phys. Rev. A} \textbf{51}, 3328 (1995)}.

\bibitem{Yuen1976-1} H. P. Yuen, \href{https://link.aps.org/doi/10.1103/PhysRevA.13.2226}{\emph{Phys. Rev. A} \textbf{13}, 2226 (1976)}.

\bibitem{Stoler1985} D. Stoler, B. E. A. Saleh, and M. C. Teich,
\href{https://doi.org/10.1080/713821735}{\emph{Opt. Acta} \textbf{32}, 345 (1985)}.

\bibitem{Lee1985} Lee and T. C., \href{https://link.aps.org/doi/10.1103/PhysRevA.13.1213}{\emph{Phys. Rev. A} \textbf{31}, 1213 (1985)}.

\bibitem{Agarwal1991} G. S. Agarwal and K. Tara, \href{https://link.aps.org/doi/10.1103/PhysRevA.43.492}{\emph{Phys. Rev. A} \textbf{43}, 492 (1991)}.

\bibitem{Lund2004} A. P. Lund, H. Jeong, T. C. Ralph, and M. S. Kim,
\href{https://link.aps.org/doi/10.1103/PhysRevA.70.020101}{\emph{Phys. Rev. A} \textbf{70}, 020101 (2004)}.

\bibitem{Wenger2004} J. Wenger, R. Tualle-Brouri, and P. Grangier,
\href{https://link.aps.org/doi/10.1103/PhysRevLett.92.153601}{\emph{Phys. Rev. Lett.} \textbf{92}, 153601 (2004)}.

\bibitem{Jing2017} Y. Li, H. Jing, and M.-S. Zhan, \href{http://dx.doi.org/10.1088/0953-4075/39/9/001}{\emph{J. Phys.B-At Mol Opt.} \textbf{39}, 2107 (2006)}.

\bibitem{Zavatta2004} A. Zavatta, S. Viciani, and M. Bellini, \href{https://doi.org/10.1126/science.1103190}{\emph{Science} \textbf{306}, p.660 (2004)}.

\bibitem{Jing2008} Y. Li, H. Jing, and M. S. Zhan, \href{https://doi.org/10.1016/j.physleta.2008.03.061}{\emph{Phys. Lett. A} \textbf{372}, 4177 (2008)}.

\bibitem{Barbieri2010} M. Barbieri, N. Spagnolo, M. G. Genoni, F.
Ferreyrol, R. Blandino, M. G. A. Paris, P. Grangier, and R. Tualle-
Brouri, \href{https://link.aps.org/doi/10.1103/PhysRevA.82.063833}{\emph{Phys. Rev. A} \textbf{82}, 063833 (2010). }

\bibitem{Dodonov1998} V. V. Dodonov, M. A. Marchiolli, Y. A. Korennoy,
V. I. Man¡¯ko, and Y. A. Moukhin, \href{https://link.aps.org/doi/10.1103/PhysRevA.58.4087}{\emph{Phys.Rev.A} \textbf{58}, 4087 (1998)}.

\bibitem{Zavatta2005} A. Zavatta, S. Viciani, and M. Bellini, \href{https://link.aps.org/doi/10.1103/PhysRevA.72.023820}{\emph{Phys. Rev. A} \textbf{72}, 023820 (2005)}.

\bibitem{Kalamidas2008} D. Kalamidas, C. C. Gerry, and A. Benmoussa,
\href{https://doi.org/10.1016/j.physleta.2007.10.089}{\emph{Phys. Lett. A} \textbf{372}, 1937 (2008)}.

\bibitem{Nuemann1955} von Neumann J, \emph{Mathematical Foundations of
Quantum Mechanics} (Princeton University Press, Princeton, NJ, 1955).

\bibitem{Aharonov1988} Y. Aharonov, D. Z. Albert, and L. Vaidman,
\href{https://link.aps.org/doi/10.1103/PhysRevLett.60.1351}{\emph{Phys. Rev. Lett.} \textbf{60}, 1351 (1988)}.

\bibitem{T2010} J. T. et al., \href{https://doi.org/10.1088/1367-2630/12/1/013023}{\emph{New J. Phys.} \textbf{12}, 013023 (2010)}.

\bibitem{Aharonov2005} {[}39{]} Y. Aharonov and D. Rohrlich, \emph{Quantum
Paradoxes- Quantum Theory for the Perplexed} (Wiley-VCH, Weinheim,
2005).

\bibitem{Hosten2008} O. Hosten and P. Kwiat, \href{https://science.sciencemag.org/content/319/5864/787}{\emph{Science} \textbf{319}, 787 (2008)}.

\bibitem{Turek2013} L. Zhou, Y. Turek, C. P. Sun, and F. Nori, \href{https://link.aps.org/doi/10.1103/PhysRevA.88.053815}{\emph{Phys. Rev. A} \textbf{88}, 053815 (2013)}.

\bibitem{Pfeifer2011} M. Pfeifer and P. Fischer, \href{http://www.opticsexpress.org/abstract.cfm?URI=oe-19-17-16508}{\emph{Opt. Express} \textbf{19}, 16508 (2011)}.

\bibitem{Dixon2009} P. B. Dixon, D. J. Starling, A. N. Jordan, and
J. C. Howell, \href{https://link.aps.org/doi/10.1103/PhysRevLett.102.173601}{\emph{Phys. Rev. Lett.} \textbf{102}, 173601 (2009)}.

\bibitem{Starling2009} D. J. Starling, P. B. Dixon, A. N. Jordan,
and J. C. Howell, \href{https://link.aps.org/doi/10.1103/PhysRevA.82.063822}{\emph{Phys. Rev. A} \textbf{80}, 041803 (2009)}.

\bibitem{Starling2010} D. J. Starling, P. B. Dixon, A. N. Jordan,
and J. C. Howell, \href{https://link.aps.org/doi/10.1103/PhysRevA.82.063822}{\emph{Phys. Rev. A} \textbf{82}, 063822 (2010)}.

\bibitem{Loaiza2014} O. S. Magaña Loaiza, M. Mirhosseini, B. Rodenburg,
and R. W. Boyd, \href{https://link.aps.org/doi/10.1103/PhysRevLett.112.200401}{\emph{Phys. Rev. Lett.} \textbf{112}, 200401 (2014)}.

\bibitem{Viza2013} G. I. Viza, J. Martínez-Rincón, G. A. Howland,
H. Frostig, I. Shomroni, B. Dayan, and J. C. Howell, \href{http://ol.osa.org/abstract.cfm?URI=ol-38-16-2949}{\emph{Opt. Lett.} \textbf{38}, 2949 (2013)}.

\bibitem{Egan2012} P. Egan and J. A. Stone, \href{http://ol.osa.org/abstract.cfm?URI=ol-37-23-4991}{\emph{Opt. Lett.} \textbf{37}, 4991 (2012)}.

\bibitem{Kofman2012} A. G. Kofman, S. Ashhab, and F. Nori, \href{https://doi.org/10.1016/j.physrep.2012.07.001}{\emph{Phys. Rep.} \textbf{520}, 43 (2012)}.

\bibitem{Dresel2014} J. Dressel, M. Malik, F. M. Miatto, A. N. Jordan,
and R. W. Boyd, \href{https://link.aps.org/doi/10.1103/RevModPhys.86.307}{\emph{Rev. Mod. Phys.} \textbf{86}, 307 (2014)}.

\bibitem{Aharonov2005-1} Y. Aharonov and A. Botero, \href{https://link.aps.org/doi/10.1103/PhysRevA.72.052111}{\emph{Phys. Rev. A} \textbf{72}, 052111 (2005)}.

\bibitem{Lorenzo2008} A. Di Lorenzo and J. C. Egues, \href{https://link.aps.org/doi/10.1103/PhysRevA.77.042108}{\emph{Phys. Rev. A} \textbf{77}, 042108 (2008)}.

\bibitem{Pan2012} A. K. Pan and A. Matzkin, \href{https://link.aps.org/doi/10.1103/PhysRevA.85.022122}{\emph{Phys. Rev. A} \textbf{85}, 022122 (2012)}.

\bibitem{Nakamura2012} K. Nakamura, A. Nishizawa, and M.-K. Fujimoto,
\href{https://link.aps.org/doi/10.1103/PhysRevA.85.012113}{\emph{Phys. Rev. A} \textbf{85}, 012113 (2012)}.

\bibitem{Lima2014} B. de Lima Bernardo, S. Azevedo, and A. Rosas,
\href{https://doi.org/10.1016/j.optcom.2014.06.008}{\emph{Opt. Commun.} \textbf{331}, 194 (2014)}.

\bibitem{Turek2015} Y. Turek, H. Kobayashi, T. Akutsu, C.-P. Sun,
and Y. Shikano,\href{https://link.aps.org/doi/10.1088/1367-2630/17/8/083029}{ \emph{New J. Phys.} \textbf{17}, 083029 (2015)}.

\bibitem{Pang2015} S. Pang and T. A. Brun, \href{https://link.aps.org/doi/10.1103/PhysRevLett.115.120401}{\emph{Phys. Rev. Lett.} \textbf{115}, 120401 (2015)}.

\bibitem{Turek2015-1} Y. Turek, W. Maimaiti, Y. Shikano, C.-P. Sun,
and M. Al-Amri, \href{https://link.aps.org/doi/10.1103/PhysRevA.92.022109}{\emph{Phys. Rev. A} \textbf{92}, 022109 (2015)}.

\bibitem{Turek2018} Y. Turek and T. Yusufu, \href{https://doi.org/10.1140/epjd/e2018-90258-8}{\emph{Eur. Phys. J. D} \textbf{72}, 202 (2018)}.

\bibitem{Kedem2010} Y. Kedem and L. Vaidman, \href{https://link.aps.org/doi/10.1103/PhysRevLett.105.230401}{\emph{Phys. Rev. Lett.} \textbf{105}, 230401 (2010)}.

\bibitem{Turek2019} Y. Turek, \href{https://arxiv.org/cits/1912.13229}{\emph{arXiv:1912.13229 [quant-ph]} (2019)}.

\bibitem{Mandel1979} L. Mandel, \href{https://link.aps.org/doi/10.1364/OL.4.000205}{\emph{Opt. Lett.} \textbf{4}, 205 (1979)}.

\bibitem{Agarwal2013} G. S. Agarwal, \emph{Quantum Optics} (Cambridge University
Press, Cambridge, England, 2013).
\end{thebibliography}
\end{document}